# Simultaneous Determination of Conductance and Thermopower of Single Molecule Junctions

Jonathan R. Widawsky[1], Pierre Darancet[2], Jeffrey B. Neaton*[2], Latha Venkataraman*[1]

[1]*Department of Applied Physics and Applied Mathematics, Columbia University, New York, NY*
[2]*Molecular Foundry, Lawrence Berkeley National Laboratory, Berkeley, CA*

AUTHOR EMAIL ADDRESS: jbneaton@lbl.gov; lv2117@columbia.edu

Abstract: We report the first concurrent determination of conductance ($G$) and thermopower ($S$) of single-molecule junctions via direct measurement of electrical and thermoelectric currents using a scanning tunneling microscope-based break-junction technique. We explore several amine-Au and pyridine-Au linked molecules that are predicted to conduct through either the highest occupied molecular orbital (HOMO) or the lowest unoccupied molecular orbital (LUMO), respectively. We find that the Seebeck coefficient is negative for pyridine-Au linked LUMO-conducting junctions and positive for amine-Au linked HOMO-conducting junctions. Within the accessible temperature gradients (<30 K), we do not observe a strong dependence of the junction Seebeck coefficient on temperature. From histograms of 1000's of junctions, we use the most probable Seebeck coefficient to determine a power factor, $GS^2$, for each junction studied, and find that $GS^2$ increases with $G$. Finally, we find that conductance and Seebeck coefficient values are in good quantitative agreement with our self-energy corrected density functional theory calculations.

Understanding transport characteristics of single metal-molecule-metal junctions is of fundamental importance to the development of functional nanoscale, organic-based devices[1]. Much work has been performed investigating the low-bias conductance of molecules attached to gold leads using a variety of chemical link groups[2-7]. However, conductance measurements provide only a partial description of electron transport through molecules, and cannot directly probe molecular energy level alignment with the electrode Fermi energy or level broadening due to electronic coupling to the leads. Thus, there is significant value in developing complimentary methods to investigate junction electronic structure and attain a more complete understanding of molecular-scale transport. In particular, molecular junction thermopower measurements can be useful in determining the dominant molecular orbital for transport and the identity of the primary charge carriers[8-11]. In a single-molecule junction, the thermopower or Seebeck coefficient $S$ determines the magnitude of the built-in potential developed across a material (or molecule) when a temperature difference $\Delta T$ is applied. With the additional presence of an external voltage bias $\Delta V$ across the junction, following Ref. 12, Equation (17), the total current $I$ in this case is simply

$$I = -G\Delta V + GS\Delta T \quad (1)$$

where $G$ is the electrical conductance. Eq. (1) applies for both bulk materials, where transport is (e.g.) diffusive[13], and in single-molecule junctions, where transport can be coherent[14]. For coherent tunneling, the conductance through a molecular junction, in the zero-bias limit, can be given by the Landauer formula[13, 14],

$$G = \frac{2e^2}{h}\mathcal{T}(E_f) \quad (2)$$

and the Seebeck coefficient by

$$S(E_f) = \frac{\pi^2 k_B^2 T}{3e\mathcal{T}(E)} \frac{\partial \mathcal{T}(E)}{\partial E}\bigg|_{E=E_f} \quad (3)$$

where $T(E)$ is the transmission function, $T$ [= $(T_1+ T_2)/2$] is the average temperature of the leads, $E_f$ is the Fermi Energy of the leads and $k_B$ is the Boltzmann constant. In this coherent tunneling limit, the current at zero external bias is then only due to a difference in temperature between the two leads, and depends on the slope of the transmission function at the Fermi energy, $E_F$. Thus, from the sign of this thermoelectric current (and therefore $S$), we can potentially deduce whether $E_F$ is closer to the highest occupied molecular orbital (HOMO) or lowest unoccupied molecular orbital (LUMO) resonance energy, assuming a simple Lorentzian-type model.

Here, we present a study of thermopower measurements for several amine-Au linked HOMO-conducting and pyridine-Au linked LUMO-conducting single-molecule junctions. In contrast to previous measurements in which thermal and electronic properties are not measured on the same junctions[8-11] or have been measured simultaneously on a junction containing a few molecules[15], we determine both the conductance, $G$, and the Seebeck coefficient, $S$, concurrently for an individual molecular junction. Conductance values are obtained by measuring the current across the gold-molecule-gold junction at an applied bias voltage of 10 mV. The Seebeck coefficient on the same junction is determined from the measured thermoelectric current through the junction held under a temperature gradient while maintaining a zero (externally-applied) bias voltage across the junction. We find that amine-terminated molecules have S>0, suggesting that the HOMO resonance is closest to $E_F$, while pyridine-terminated molecules have S<0, indicating that the LUMO resonance is closest. We also find that a diphenylphosphine-terminated alkane has an $S$ near zero, indicating that $E_F$ is probably very close to the middle of the HOMO-LUMO gap. We compare our measurements with first principles transport calculations that are based on standard density functional theory (DFT) that is extended to incorporate self-energy corrections[16, 17], and find quantitative agreement with both conductance values and Seebeck coefficients. Our calculations confirm the sign of *S* in the case of each junction, and its expected relationship with the frontier orbital closest to $E_F$, but also reveal a complex transmission for amine-linked junctions, with implications for the relationship between *S* and *G*. In particular, for amine-linked junctions, *T(E)* is non-Lorentzian and thus *S* varies significantly more than *G* from junction to junction.

Single molecule junctions are created using the scanning tunneling microscope-based break junction technique (STM-BJ), in which a sharp gold tip is brought in and out of contact with a gold substrate in an environment of a target molecule[3, 18]. The molecules used in this study are 4,4'-diaminostilbene (**1**), bis-(4-aminophenyl)acetylene (**2**), 1,5-bis(diphenylphosphino)pentane (**3**), 4,4'-bipyridine (**4**) and 1,2-di(4-pyridyl)ethylene (**5**). The target molecules are deposited onto the STM substrate by thermal evaporation under ambient conditions (except for (**3**) which is deposited from an acetone solution) and thus measurements are not carried out in solvent. A thermal gradient is applied using a Peltier heater to controllably heat the substrate to temperatures ranging from room temperature to 60°C while maintaining the tip close to room temperature, and the thermoelectric current through these junctions is measured at zero applied bias. The set-up is allowed to come to thermal equilibrium for about one hour at each temperature before measurements are continued. In all measurements reported here, the temperature difference (Δ*T*) between the tip and substrate is set at approximately one of three values: 0K, 14K, 27K. In order minimize unaccountable thermoelectric voltages across the leads, a pure gold wire is used to apply the voltage to the hot substrate and is also connected to the cold side of the peltier. A schematic of the circuitry as well as the STM layout is shown in Figure 1A.

Conductance and thermoelectric current measurements are carried out using a modified version of the STM-BJ technique[19]. Briefly, the Au STM tip is brought in contact with the heated Au substrate while applying a bias voltage of 10 mV until a conductance greater than 5 $G_0$ is measured. The tip is first retracted from the substrate by 2.4 nm at a speed of 15.8 nm/s, then held fixed for 50 ms, and finally withdrawn an additional 0.8 nm as illustrated by the piezo ramp shown in Figure 1B. During this ramp, the current through the junction and the voltage across the junction is continuously measured at a 40 kHz acquisition rate. The applied bias is set to zero during the middle 25 ms of the 50 ms period when the tip/substrate distance is fixed (Figure 1C). For every molecule, and tip/sample pair, over 3600 measurements are collected at each of the three $\Delta T$'s listed above. For each measurement, the data during the 50 ms hold period are analyzed further, as described in detail in the SI. Briefly, the junction conductance during the first and last 12.5 ms of the "hold" period is determined. If both values are found to be within a molecule dependent range as determined from a conductance histogram, the trace is selected, for this indicates that a molecular junction is sustained during the entire 50 ms "hold" period. Conductance histograms for **4** and **5**, show two peaks due to two different binding geometries[17]. Here, we analyze these molecular junctions in the high conductance configuration. The analysis of **4** in the low-conducting configuration is included in the SI. Typically, about 10-20% of the measured traces are selected for analysis, since only these have a molecule bridging the tip and substrate during the "hold" period of the ramp. A single selected measurement for **1** is shown in Figure 1B, where measured current is in red and measured voltage is in blue. The junction thermoelectric current is determined by averaging the measured junction current during the middle 25 ms of the "hold" period and the junction conductance is determined by averaging the conductance during the first and last 12.5 ms of the "hold" period. This allows a determination of junction conductance and thermoelectric current for each individual junction formed, and thus allows a simultaneous determination of G and S for each junction.

In Figure 2, we show histograms of conductance values and average thermoelectric currents for **1** and **4** determined on a trace by trace basis for measurements at three different $\Delta T$ values. Conductance and thermoelectric current distributions for other molecules studied are shown in the SI. We see that with a $\Delta T$ of 0 K, the thermoelectric current histogram is narrow and centered about zero for both molecules, implying that on average, no current flows without an applied temperature difference between the tip and substrate, consistent with the expression for current in Equation 1. For molecule **1**, we find that with a finite $\Delta T$, the thermoelectric current (from the tip to the substrate) is negative while for **4** we measure a positive thermoelectric current. We also see that the peak of the thermoelectric current distributions for **1** (**4**) shifts to lower (higher) value with increasing $\Delta T$, thus the magnitude of the thermoelectric current increases with increasing $\Delta T$.

The Seebeck coefficient for the entire system, $S_{Measured}$, is determined by first substituting, in Equation (1), the conductance and thermoelectric currents determined for each measured trace ($S_{Measured} = I/G\Delta T$). Since the system includes a section of gold wire which is maintained under the opposite thermal gradient ($-\Delta T$), the Seebeck coefficient of the Au-molecule-Au junction is given by $S_{Junction} = S_{Au} - S_{Measured}$, where we use $S_{Au}$ = 2 µV/K[20]. In Figure 3, we show the distribution of molecular Seebeck coefficients (when the molecules are attached to Au electrodes) determined for all five compounds studied by including measurements at all $\Delta T$s. These distributions are fit to a Gaussian and the peak positions are given in Table 1, along with measured conductance values. We use the most probable molecular junction conductance and Seebeck coefficients to determine a power factor, $GS^2$, for these systems, which are also given in Table 1. We see first that the amine-terminated molecules (**1** and **2**) have a positive Seebeck coefficients indicating HOMO conductance, while the **4** and **5** have a negative Seebeck coefficient (LUMO-conducting). For molecule **3**, although we measure a small positive $S_{Junction}$, the magnitude is small enough to conclude that for this alkane, $E_F$ is very close to mid-gap.

To understand these measurements, we use first-principles calculations with a self-energy corrected, parameter-free scattering-state approach based on density functional theory (DFT)[17, 21, 22] to determine both the linear response conductance (Eq. 2) and the Seebeck coefficient (Eq. 3). Eq. (3) assumes that $T(E)$ varies smoothly for $|E - E_F| < k_BT$, and that $\Delta T$ is small compared to $T$[14, 23]. Both assumptions hold for the systems studied here ($|\Delta T| < 30$ K). Moreover, since the measured thermoelectric current is found to be approximately linear with $\Delta T$ for the small values of $\Delta T$ in the experiments, we expect the steady-state scattering formalism will also be valid[24].

We model the electrodes using two Au(111) slabs with 7 layers of gold for the leads, using a 4×4 unit cell. To reduce computational burden for molecule **3**, we replace the phenyl groups with methyl groups, a simplification that has been shown not to affect conductance experimentally[25]. Previous work established that for both amine- and pyridine-Au linked molecular junctions that the amine or pyridine group binds selectively to undercoordinated atop Au sites[21, 22]. Accordingly, we use two different undercoordinated binding site motifs, consisting of either a trimer of gold or an adatom to represent the tip to which the molecules bind[16, 21]. All junction geometries are fully relaxed within DFT-GGA (PBE) using SIESTA[26]. Transmission functions are calculated using a self-energy corrected scattering-states approach, "DFT+Σ", to the Landauer formula as implemented in the Scarlet code[27] (see SI for details).

Junction structures for molecules **1** and **4** are shown in Figure 4. To calculate the junction transmission, we augment the Kohn-Sham excitation energies with a model self-energy correction that has consistently led to quantitative agreement for both conductance[17, 21, 22] and Seebeck coefficient[16]. Specifically, we correct the gas-phase gap with a $\Delta$SCF[28] calculation, and correct for the lack of static non-local correlation effects through an electrostatic "image charge" model, following prior work[17]. $G$ and $S$ are then determined from the transmission and its derivative at $E_F$, via Eq. 2 and 3. Numerical evaluation of the derivative of $T(E)$ generally requires a very fine $k_{||}$-point sampling. To minimize sampling errors, we fit $T(E)$ around $E_F$ with a smooth function, and take its derivative analytically. Comparing these two approaches for one junction, 1,4-benzenediamine-Au[16], we find a less than 5% difference in $S$ obtained from numerically differentiating $T(E)$ on a 24×24 $k_{||}$-grid and fitting a $T(E)$ calculated on an 8×8 $k_{||}$-grid. For all the results presented here, $T(E)$ is computed on a 16x16 $k_{||}$-grid, with S determined from the analytic derivative of a fit to the transmission function around $E_F$.

The transmission curves for **1** and **4** are shown in Figure 4 (other molecules are shown in SI Figure S7) and the calculated values for *G* and $S_{Junction}$ are reported in Table 1. We find that for amine-linked junctions, $T(E_F)$ originates with a HOMO-derived peak, and a weakly-transmitting feature formed from a hybridization of Au-d and N-lone pair states, resulting in a positive *S*. The calculated *G* are within 15% of the experiments, and do not vary much with the coordination of the Au binding site, in agreement with past work showing that junction geometry does not affect conductance significantly[21, 29]. For the pyridine-linked junctions, transmission at Fermi results from a LUMO derived peak, hence *S* is negative. The calculated *G* is within a factor of 3 of the experiment but varies significantly with the Au binding site coordination, again in agreement with past work[30]. The calculated Seebeck coefficients for both series are within a factor of 2 of the experimental values.

For the amine-linked junctions, hybridization between the HOMO resonance and Au d-states ~1.8 eV below $E_F$ (cf. arrow in Figure 4A) results in an appreciable energy-dependent coupling, and the single Lorentzian model breaks down. Increasing the binding site coordination significantly alters the lineshape and the Seebeck coefficients (15% variation for molecule **1**), but results in just a modest rise in the density of states at $E_F$ and small changes in the conductance (3% variation). Interestingly, *S* is more sensitive to HOMO-Au 5d hybridization features in $T(E)$ than *G*. Because these features may be underestimated by approximations associated with DFT+Σ, deviations from experiment may be somewhat greater for *S* than for *G* for HOMO-conducting amine-Au junctions. For molecules **1** and **2,** the simultaneous measurement of *G* and *S* allows an assessment of the efficacy of Lorentzian models. If the transmission function has a simple Lorentzian form for a molecule symmetrically coupled to both electrodes[8, 9], *S* and *G* determine the resonance energy relative to $E_F$. However, for the amine-Au linked junctions studied here, a single-Lorentzian model (see SI) significantly underestimates the resonance energy. For molecule **1**, the single-Lorentzian model would place the HOMO at −1.1 eV while our DFT+Σ calculations show that HOMO is around −2.3 eV. Thus the non-Lorentzian behavior of amine-linked junctions observed here, and seen with some other linkers[7] could allow for the tuning of their *S* without greatly affecting *G* through the position of the d-states, for example using transition metal contacts with d-states closer to $E_F$.

For pyridine-linked junctions, the calculated DFT+Σ transmission function has a Lorentzian form, with a prominent resonance with LUMO character. Increasing the binding site coordination pushes the LUMO resonance away from $E_F$. In a single Lorentzian model in the weak-coupling limit, $(\Gamma/\Delta E)^2 \ll 1$ (where $\Gamma$ is an energy-independent coupling or injection rate, and $\Delta E$ the difference between the LUMO resonance energy and $E_F$), and the Seebeck coefficient varies more slowly with $\Delta E$ (as $1/\Delta E$) than the conductance ($1/\Delta E^2$). This is indeed what we find in our calculations. For molecule **4,** *S* has a +/-5% variation for the different binding sites, while the conductance changes by +/-25%. This Lorentzian-like behavior is further validated by the estimate of the resonance positions, in close agreement with first-principles calculations (1.53 eV determined using the experimental *S* and *G* compared with 1.47 eV from our calculations of molecule **4**).

In conclusion, we have demonstrated that we can determine the conductance and thermoelectric current concurrently through single-molecule junctions. The thermoelectric currents are used to determine a Seebeck coefficient for each junction. We find that amine-terminated molecular junctions have a positive Seebeck coefficient in agreement with calculations that show that the HOMO is the molecular resonance that is closest to $E_F$. In contrast, pyridine-terminated molecular junctions have a negative Seebeck coefficient and conduct through the LUMO. These experimental results are in good quantitative agreement with those from self-energy corrected DFT calculations, which also reveal a complex, non-Lorentzian form for transmission for amine-linked junctions.

**Acknowledgements:** We thank M. S. Hybertsen, H. J. Choi, and S. Y. Quek for discussions and S. Berkley for help with measurements. This work was supported in part by the EFRC program of the U.S. Department of Energy (DOE) under Award No. DE-SC0001085 and the ACS-PRF program. L.V. thanks the Packard Foundation for support. Portions of this work were performed at the Molecular Foundry and within the Helios Solar Energy Research Center, both were supported by the Office of Basic Energy Sciences of the U.S. Department of Energy under Contract No. DE-AC02-05CH11231. We acknowledge NERSC for computing resources.

Table 1

| Molecule | $G_{EXP}$ ($10^{-3}$ $G_0$) | $S_{EXP}$ (µV/K) | $[GS^2]_{EXP}$ ($10^{-18}$ W/K²) | $G_{DFT+\Sigma}$ ($10^{-3}$ $G_0$) | $S_{DFT+\Sigma}$ (µV/K) |
|---|---|---|---|---|---|
| 1 | 0.63 | 13.0 (7.0) | 8.25 | 0.58 | 5.89 |
| 2 | 0.57 | 9.7 (6.1) | 4.16 | 0.62 | 4.71 |
| 3 | 0.39 | 1.1 (4.1) | 0.037 | 0.70 | 0.33 |
| 4 | 0.68 | -9.5 (4.3) | 4.76 | 0.2 | -7.88 |
| 5 | 0.24 | -12.3 (9.1) | 2.81 | 0.07 | -12.11 |

**Table Caption**

Table 1: List of molecular conductance, *G*, and Seebeck coefficient, *S* (with the HWHMs included in parentheses), determined experimentally and theoretically. Also shows the experimentally determined power factor, $GS^2$.

**Figure Captions**

**Figure 1: (a)** Top panel: Simplified diagram illustrating measurement of thermoelectric current ($I_T$). Bottom panel: Schematic of the STM-BJ set-up. **(b)** Top panel: Piezo ramp used, including a "hold" portion between 150 and 200 ms. Middle panel: External applied voltage across the leads which drops to zero during the center of the "hold" portion. Bottom panel: Sample trace for molecule **1**. The measured current is shown in red and the voltage measured across the junction is shown in blue. Note: The voltage is applied across the junction in series with a 10 kΩ resistor.

**Figure 2: (a)** Average measured conductance histograms for molecules **1** (top) and **4** (bottom), for the three ΔT's (ΔT=0 K, green; ΔT=14 K, blue; and ΔT=27 K, red). **(b)** Average thermoelectric current histograms for molecules **1** (top) and **4** (bottom), for the three ΔT's (ΔT=0 K, green; ΔT=14 K, blue; and

ΔT=27 K, red). For **1**, the thermoelectric current shifts left with increasing ΔT, while for **2**, the thermoelectric current shifts right.

**Figure 3:** Histograms of Seebeck coefficient for all molecules **1**–**5**. Histograms are fit with Gaussians (red). **1** and **2** exhibit positive *S* while **4** and **5** exhibit negative *S*. The Seebeck coefficient for **3** is close to zero.

**Figure 4:** Upper panel: The optimized geometries for junctions with molecules **1 (a)** and **4 (b)**. Lower panel: Transmission curves shown on a log scale for both molecules calculated using DFT+Σ. Arrow indicates the position of the Au-d states. Insets: Transmission curves around $E_F$ on a linear scale.

## Figure 1

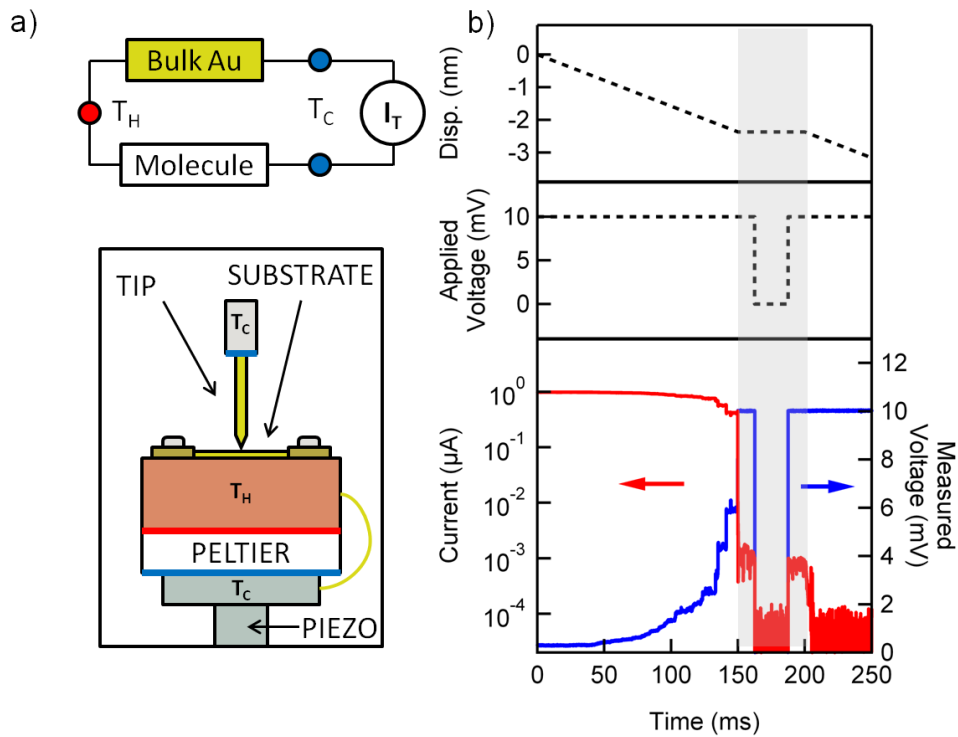

## Figure 2

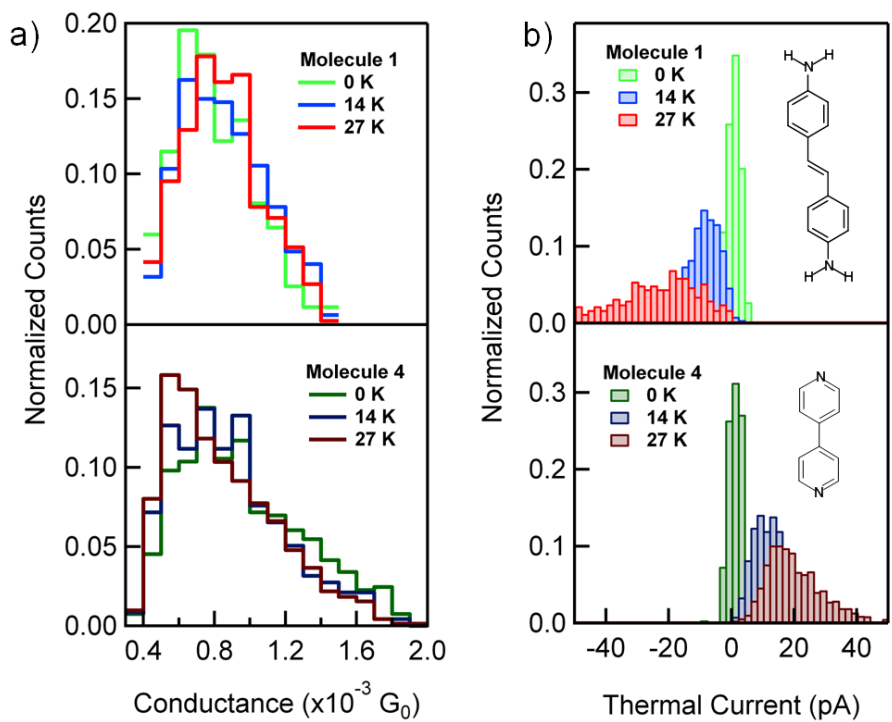

Figure 3

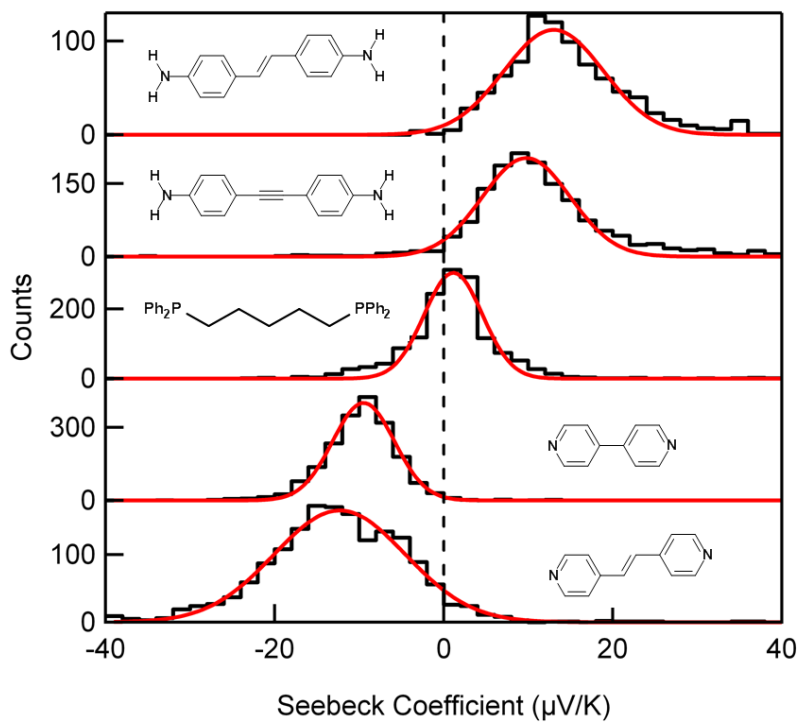

Figure 4

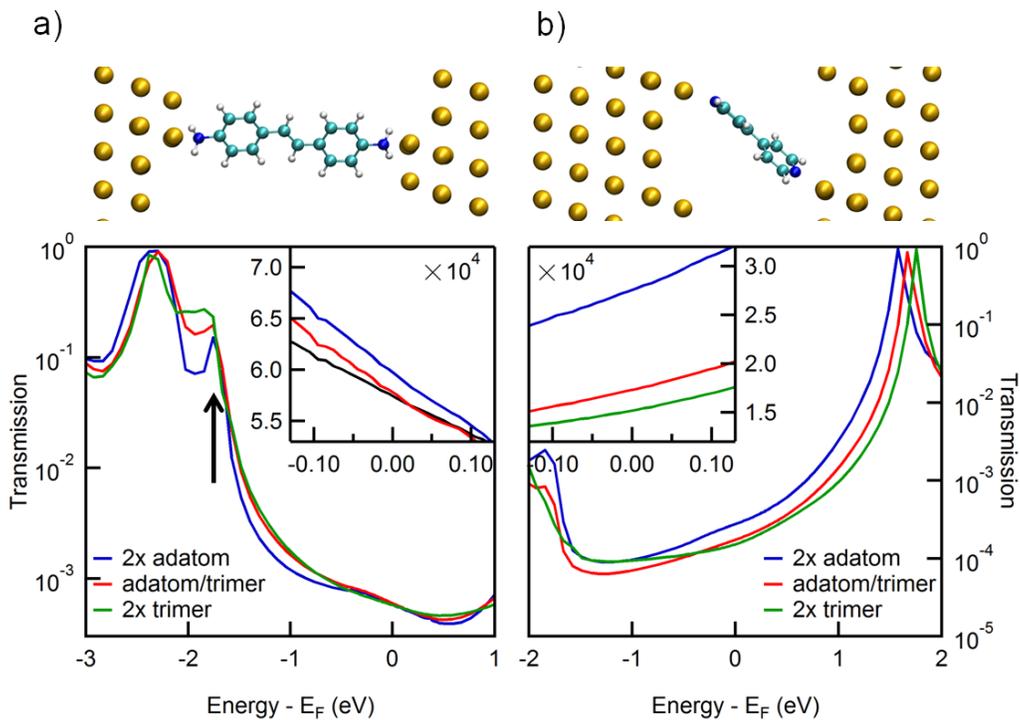